%% file: main.tex
\documentclass[aps,a4paper,pre,preprint,floatfix,groupedaddress,tightenlines]{revtex4-1}
\pdfoutput=1

\usepackage{graphicx,mathrsfs,amsmath,amssymb,amsopn,mathbbol,bm,qmech-revtex}
\usepackage{color}
\usepackage[verbose]{placeins}

\begin{document}
\title{A Monte Carlo framework for noncontinuous interactions between particles and classical fields}

\author{Christian Wesp}
\email[]{cwesp@th.physik.uni-frankfurt.de}

\affiliation{Institut f\"ur theoretische Physik, Goethe-Universit\"at
  Frankfurt am Main, Max-von-Laue-Stra{\ss}e 1, 60438 Frankfurt, Germany}

\author{Hendrik van Hees}

\affiliation{Institut f\"ur theoretische Physik, Goethe-Universit\"at
  Frankfurt am Main, Max-von-Laue-Stra{\ss}e 1, 60438 Frankfurt, Germany}

\author{Alex Meistrenko}

\affiliation{Institut f\"ur theoretische Physik, Goethe-Universit\"at
  Frankfurt am Main, Max-von-Laue-Stra{\ss}e 1, 60438 Frankfurt, Germany}

\author{Carsten Greiner}

\affiliation{Institut f\"ur theoretische Physik, Goethe-Universit\"at
  Frankfurt am Main, Max-von-Laue-Stra{\ss}e 1, 60438 Frankfurt, Germany}

\date{\today}

\begin{abstract}
  Particles and fields are standard components in numerical calculations
  like transport simulations in nuclear physics and have well understood
  dynamics. Still, a common problem is the interaction between particles
  and fields due to their different formal description. Particle
  interactions are discrete, point-like events while field dynamics is
  described with continuous partial-differential equations of motion. A
  workaround is the use of effective theories like the Langevin equation
  with the drawback of energy conservation violation. We present a
  method, which allows to model noncontinuous interactions between
  particles and scalar fields, allowing us to simulate scattering-like
  interactions which exchange discrete ``quanta'' of energy and momentum
  between fields and particles while obeying energy and momentum
  conservation and allowing control over interaction strengths and
  times. In this paper we apply this method to different model systems,
  starting with a simple harmonic oscillator, which is damped by losing
  discrete energy quanta. The second and third system consists of an
  oscillator and a one-dimensional field, which are damped via discrete
  energy loss and are coupled to a stochastic force, leading to
  equilibrium states which correspond to statistical Langevin-like
  systems. The last example is a scalar field in (1+3) space-time
  dimensions, which is coupled to a microcanonical ensemble of particles
  by incorporating particle production and annihilation
  processes. Obeying the detailed-balance principle, the system
  equilibrates to thermal and chemical equilibrium with dynamical
  fluctuations on the fields, generated dynamically by the discrete
  interactions.
\end{abstract}


\maketitle

\section{\label{sec:Introduction}Introduction}

Monte-Carlo simulations are a mature tool in theoretical, experimental
and computational physics to simulate a broad variety of problems. By
randomly sampling a possible system state of a complex system ensemble,
they can be applied to problems which technically do not have an
immediate probabilistic interpretation \cite{carter1975particle}. In
nuclear physics, Monte-Carlo simulations have been used to integrate the
discretized action in Lattice QCD problems \cite{rothe2012lattice}; in
transport simulations they are used to sample collision integrals and
phase-space distributions of particles \cite{friman2011cbm}. Beside
approaches with simple Gaussian (or white) random processes, complex and
microscopic interaction kernels, based on physical properties, can be
employed with help of the Boltzmann-equation \cite{Xu:2004mz}.
Hydrodynamics is another, very successful approach to the simulation of
physical systems by propagating conserved quantities on a mesoscopic
scale. Hydrodynamics is not directly interpreted as a classical field
theory. However, its mathematical description resembles the one of
fields, for example fields for the energy density and pressure
(energy-momentum tensor). Both classical field theories and
hydrodynamics are described by a set of deterministic, coupled partial
differential equations and are generally not in the scope of Monte-Carlo
methods. To implement features of noise and fluctuations, different
approaches have been developed to couple random processes to the
equations of motion to mimic microscopic fluctuating forces
\cite{ermak1978brownian, fixman1986construction}. This implements random
processes and fluctuations in the systems but has the disadvantage of
having only a statistical and long-term control over the processes.
Additionally, these methods imply continuous interactions and
dissipation between fields and their external forces or sources. Trying
to apply those methods to particle-field interactions will bring
problems as processes like particle creation and annihilation are not
continuous processes.

In this paper we present a new approach which allows interactions
between particles and fields without the need of effective random
processes. The underlying physical interactions and processes can be
modeled and simulated with Monte-Carlo methods; their impact on the
system change are then propagated back to the fields and particles. This
is achieved by precisely controlling momentum and energy conservation at
all times of the simulation. Various examples are given in this paper,
including the simulation of a thermal box with inelastic interactions
between particles and fields.

This paper is organized as follows. In the first section, we summarize
current methods to simulate interactions between fields and particles
employing Monte-Carlo like methods and discuss their advantages and
disadvantages. In section \ref{MyMethod} we introduce the framework of
our new method to model and simulate noncontinuous interactions between
fields and particles. Sec.\ \ref{sec:TestCalculations} shows different
examples for the implementation of our method in various test
scenarios. Applications to various other physical fields and disciplines
are discussed in \ref{sec:Applications}. Finally, we conclude with a
short summary.

\section{Approaches for particle-field interactions\label{Classical approaches}}
\subsection{Vlasov-Approach}

The classical Boltzmann-equation is used to describe the transport and
interactions of particles in the dilute-gas limit
\cite{reichl1980modern}. The particles are described by a continuous
distribution function which is propagated by a set of
integro-differential equations. To describe charged particles in a
plasma, Vlasov used the collisionless Boltzmann equation and coupled the
particles to self-consistent field potentials \cite{0038-5670-10-6-R01,
  dupree1966perturbation, birdsall2004plasma} resulting in the famous
(relativistic) Vlasov-equation,
\begin{equation}\label{eq:Vlasov}
  \frac{\dd f}{\dd t} = 
  \frac{\partial f}{\partial t} 
  + \frac { \bvec{p}}{E} \cdot  \frac{\partial f}{\partial \bvec{x}} 
  + \bvec{F}(\bvec{x},\bvec{p}, t) \cdot
  \frac{\partial f}{\partial \bvec{p}} = 0 \ .
\end{equation}
Eq. (\ref{eq:Vlasov}) describes the propagation of a distribution
function in position and momentum space, the interesting part is the
interplay between the change of momenta by the external force $\bvec F$
and the propagation of particles by their velocity $\bvec p/E$.  The
external force $\bvec F$ can be given by an external potential and/or a
mean-field interaction with the surrounding particles.

A possible way to couple a field $\phi$ with a particle distribution
function $f(\bvec{x},\bvec{p},t)$ is to employ a Yukawa interaction
\cite{raey}. Particles are accelerated by the gradients of the fields
\begin{equation}
  \bvec{F} = - \nabla_{\bvec x} E(\phi) \ ,
\end{equation}
corresponding to a mean field mean field coupling which cannot
thermalize the system as it leads to reversible dynamics. By employing
interactions with higher-order loops \cite{Berges2001369,
  PhysRevD.55.6471} or by coupling the mean field to particles with
collision kernels, thermalization can be recovered but on large time
scales. Additionally, the interactions within the Vlasov equation are
very soft and do not include inelastic processes like particle
production and annihilation. Nevertheless these processes can play an
important role in some applications. The linear $\sigma$-model, for
example, shows a shift in thermodynamic properties when particle-number
changing processes are neglected, and in dynamical calculations the
chiral phase transition is lost if the field can not dissipate energy by
particle production and annihilation upon a temperature change
\cite{WespPhDThesis}.

\subsection{Momentum Space Approach}

A common approach to approximate quantum field systems is to assume
spatial isotropy and reduce the phase-space distribution of the physical
problem to functions in momentum space. This ansatz has been very
successful and answered many questions regarding non-equilibrium
phenomena and thermalization \cite{PhysRevD.64.105010, Berges:2002wr,
  PhysRevD.69.025006}. Additionally, in quantum field theory
interactions and their propagators are often well-defined in momentum
space, leading to transition probabilities for single Fourier modes in
terms of scattering-matrix elements,
$S_{fi}=\matrixe{\bvec{p}_1' \bvec{p}_2' \ldots}{S}{\bvec{p}_1
  \bvec{p}_2 \ldots} $.
These probabilities can be treated perturbatively, like in quantum
electrodynamics (QED). In numerical simulations, such dynamics can be
applied to a classical field, and the impact on the system by changing
single Fourier modes is easily calculated through the sum of the modes'
energies. This method works well for systems in momentum space with the
assumption of spatial isotropy. In the case of spatial anisotropy or if
the system propagation has to be described in position space and only
the interactions are performed in momentum space, then this method
implies a violation of causality. The change of a single mode in
momentum space changes the field distribution instantaneously in
position space over the whole volume, resulting in superluminal signal
propagation. These artifacts render this method unsuitable for
simulations which rely on position-space calculations.

\subsection{Langevin Equation}

The Langevin or Ito-Langevin equation \cite{ajp/65/11/10.1119/1.18725}
is a stochastic differential equation, used to describe systems with two
different scales. The ``macroscopic'' long-range and slow-timescale part
is described by deterministic equations of motion. Additionally it is
coupled to the ``microscopic'' small scales, which are described by
short-ranged fast random processes.  For a heavy particle in a heat
bath, the original Langevin equations reads
\begin{equation}\label{eq:PlainLangevin}
 m \ddot x(t) = - \gamma \dot x(t) + F_{\mathrm{ext}}(x,t) + \xi(t)
\end{equation}
with an external force $F_{\mathrm{ext}}$, the linear damping
coefficient $\gamma$, and a stochastic force $\xi$.  Often a simple
Gaussian white-noise process is assumed for the fluctuating force, which
leads to an average energy flux from and into the bulk system, while the
damping is the dissipative average part with the back reaction of the
medium to the energy-momentum exchange neglected.

In nuclear physics, the Langevin Equation has been applied on top of the
Boltzmann equation to include fluctuations in the system
\cite{Bonasera19941,Randrup1990339,Chomaz1991340, Greiner1998328}.  By
dividing dynamics of a scalar quantum field in a hard and a soft part, a
stochastic description of the system can be employed which resembles a
Langevin equation \cite{PhysRevD.55.1026}. The Langevin equation can be
used to investigate fluctuations in the linear $\sigma$-model
\cite{PhysRevLett.79.3138} or with similar methods to investigate
disoriented chiral condensates \cite{PhysRevD.62.036012}. Using the
influence formalism, classical equations for the $O(N)$ fields at
presence of a heat bath can be derived, when a stochastic interpretation
is employed \cite{PhysRevC.58.2331}.  In \cite{Nahrgang2012109,
  Nahrgang:2011mg, herold2013chiral} the Langevin equation has then been
employed to phenomenologically model a stochastic coupling of a locally
equilibrated (hydrodynamical) particle bath and a classical field within
a linear $\sigma$-model. This coupling allows an effective
thermalization of the mean field.

However, the Langevin equation has some drawbacks, the dissipation of
the equation (\ref{eq:PlainLangevin}) due to the friction term,
$\gamma \ {\dd \phi(t)}/{\dd t}$ is a continuous process. This is a
natural assumption for continuous systems like fields or waves and a
reasonable approximation for systems with a clear separation of scales,
like in the classical example of a heavy particle in a bath of small
ones. However, many processes are discrete and occur as single
events. The same problem holds for the random force, $\xi$, which acts
continuously and changes its value with every time step in numerical
implementations. Because of the random nature of this process, the exact
amount of exchanged energy can only be controlled in a statistical
manner, and the back reaction of the bulk medium is neglected. In most
implementations, the random force $\xi(t)$ is modeled by Gaussian white
noise without a memory kernel. Using a more sophisticated ansatz with
memory kernel, the random force can be extended to a non-Markovian
stochastic process with colored noise \cite{schmidt2014simulation}.

Before we discuss the relation between momentum and energy dissipation
within a Langevin equation, we have to define them for a field
$\phi$. For a general Klein-Gordon equation,
\begin{equation}
  \partial_\mu \partial^\mu \phi + m^2 \phi  + \frac{\partial U}{\partial \phi} = 0 \ ,
\end{equation}
the following conserved quantities can be defined
\cite{Yang06numericalstudies,whitham2011linear}:
\begin{alignat}{2}
\label{theory:fieldParticle:fieldEnergy}
\begin{split}
  E &= \int_V \dd^3 x \, \epsilon(\bvec{x}) \\
  &= \int_V \dd^3 x \, \left [ \frac{1}{2} \dot \phi^2 + \frac{1}{2} \left
      (\vec \nabla \phi \right)^2 + U(\phi) \right ],
\end{split}
\\
\label{theory:fieldParticle:fieldMomentum}
\bvec{P} &= \int_V \dd^3 x \, \bvec{\Pi} (\bvec x) = \int_V \dd^3 x \, \dot \phi
\vec{\nabla} \phi,
\end{alignat}
where $E$ denotes the total field energy, and $P$ is the total
momentum. For any positive-definite potential $U$, the relation
\begin{equation}
P \leq E
\end{equation}
holds. The dissipative part of the Langevin equation for the field
$\gamma \partial_t \phi$ damps both the energy
(\ref{theory:fieldParticle:fieldEnergy}) and the momentum
(\ref{theory:fieldParticle:fieldMomentum}). For a potential-free wave
equation with damping,
\begin{equation}
\partial_t^2 \phi(t,x) + \gamma \dot \phi(t,x) =\nabla^2 \phi(t,x) \ ,
\end{equation}
the ratio of $P(t)/E(t)$ is non-linear in time because both quantities
are non-linear operators while $\dot \phi$ is linear. This results in
different damping rates for $E$ and $P$ and complicates any attempt to
couple particles and fields through inelastic interactions within an
effective model.

Another problem arises with the continuous nature of the dissipative
term in the Langevin equation. For a continuous process, quantities like
energy transfer can be calculated by integrating over a time interval,
but this leads to a continuous value which can not be interpreted by an
integer number of events. In contrast, singular events like
particle-pair production and annihilation can be counted, and rates are
defined in a statistical manner. This becomes a problem when one tries
to couple a scalar field to an ensemble of particles with interactions
given by pair production and annihilation. Energy loss of the scalar
field leads to energy gain in the particle ensemble and vice versa. Such
an ansatz is used in the famous and successful cosmological inflation
model \cite{PhysRevLett.73.3195}, in which particles are created by the
energy loss of the oscillating scalar field, $\Phi$. Particle production
is described by rate equations, which are derived from the fields'
equations of motion. Trying to simulate such a process with finite
ensembles of particles leads to different problems. The energy loss
within a time step $\Delta t$ can be calculated from the fields and
mapped to a certain number of created particles. The energy of a
discrete number of created particles will, however, never match exactly
with the continuous loss rate. Additionally, the physical process of
pair production will depend on the simulation time-step size, and for
$\Delta t \to 0$ a mapping between the continuous dissipation and the
noncontinuous particle creation will not be possible anymore. Another
problem is the fact that the random force $\xi(t)$ changes its value at
every point in time, both for white and colored noise.  Trying to couple
this behavior to pair production and annihilation leads to the same
problem as the microscopic processes will depend on the time step.

In summary, the Langevin equation is a very good choice for an effective
description of a system with two separate time scales. However, a
microscopic modeling of the interaction processes is complicated by the
continuous nature of the Langevin process. In the next section we will
present how to potentially solve these problems by a noncontinuous
approach.

\section{noncontinuous interactions between fields and particles}
\label{MyMethod}

This work is originally motivated by the need to describe inelastic and
discrete interactions between fields and particles. The process of
creating particles and anti-particles from field excitations and vice
versa,
\begin{equation}
\bar{q} + q \leftrightarrow \phi \ ,
\end{equation}
turns out to be very important for a consistent description of chemical
and thermal equilibration in the linear $\sigma$-model and in our
transport simulation DSLAM ({\bf D}ynamical {\bf S}imulation of a {\bf
  L}inear Sigm{\bf A} {\bf M}odel) \cite{WespPhDThesis}. The particles
are modeled with the test-particle ansatz, and the fields are
represented by classical scalar fields. Any interactions beyond
mean-field couplings have to be discrete, because energy and momentum
can only be transferred in terms of whole particle pairs; we therefore
have developed a particle-field method which allows such noncontinuous
interactions. The method is split up in two parts: a framework for
transferring an exact amount of energy and momentum from and to a field
and a field theoretical calculation for taking possible interactions
between fields and particles into account in a microscopic way. The
second method enables us to develop Monte-Carlo models for the
interaction probability of a given process, and the first method allows
us to realize the interactions in a simulation.

\subsection{\label{sec:EnergyTransfer}Discrete energy and momentum transfers from and to a field}

Particles and fields are described in a very different manner. While
scalar fields define a single quantity as function of time and position
in continuous space (or discrete space on a grid), particles are
characterized by their position and momentum. For particles, this sums
up to three position coordinates, the energy and three-momentum values
at every point in time. Fields are described by continuous functions or
$N^3$ values on a three-dimensional $N$-sized grid for the field
$\phi(\bvec{x}, t)$ and its time derivative, $\dot \phi(\bvec{x}, t)$.

To link these very different descriptions, common mathematical and
physical properties have to be found. The most simple approach is to use
energy and momentum. For particles, energy and momentum are directly
given by their momentum four-vector. For a field we can employ the
already discussed relations (\ref{theory:fieldParticle:fieldEnergy}) and
(\ref{theory:fieldParticle:fieldMomentum}). A discrete interaction now
maps to a discrete change of energy and momentum at a given position
$\bvec{x}$ and time $t_k$. The field $\phi(\bvec{x}, t_k)$ is propagated
with its undisturbed equations of motion and changed due to an
interaction by a kick $\delta \phi(\bvec{x}, t_k)$ which changes the
energy and momentum by the desired amount $\Delta E$ and
$\Delta \bvec{P}$. This leads to relations of four coupled non-linear
equations,
\begin{alignat}{2}
\label{eq:theory:FieldManipulationEnergy}
  \Delta E(t_k) &= E\left[\phi(\bvec{x}, t_k) + \delta \phi(\bvec{x}, t_k) \right] -
  E\left[\phi(\bvec{x}, t_k)  \right],\\
\Delta \bvec{P}(t_k) &= \bvec{P}\left[\phi(\bvec{x}, t_k) + \delta
  \phi(\bvec{x}, t_k) \right] - \bvec{P} \left[\phi(\bvec{x}, t_k)
\right]. \label{eq:theory:FieldManipulationMomentum}
\end{alignat}
In general (\ref{eq:theory:FieldManipulationEnergy}) and
(\ref{eq:theory:FieldManipulationMomentum}) have to be solved with a
numerical non-linear equation solver. For $\Delta E < 0$ energy will be
taken out of the field, $\Delta E > 0 $ will add this amount energy (and
analogously for the momentum).

Without further constraints, Eqs.\
(\ref{eq:theory:FieldManipulationEnergy}) and
(\ref{eq:theory:FieldManipulationMomentum}) have either no solutions or
infinitely many. To define and find unique solutions, the disturbance
kick $\delta \phi(\bvec{x}, t_k)$ has to be parameterized. In general,
$\delta \phi(\bvec{x}, t_k)$ must have a finite support to keep
causality. Furthermore it can not be a point like disturbance as this
will cause instabilities in the field equations as well as numerical
problems \cite{WespPhDThesis}. Therefore, the parameterization should be
as smooth as possible to avoid shocks and numerical artifacts on the
scalar field.

A useful and robust parameterization is a three-dimensional, moving
Gaussian wave packet,
 \begin{equation}\label{eq:parameterization:gauss3D}
   \delta \phi(\bvec{x}, \bvec{v}) = A_0 \left . \prod_i^3 \exp \left [
       -\frac{(x_i - v_i \tilde t)^2}{2\sigma_i^2}\right ] \right |_{\tilde t \to 0},
\end{equation}
where $\bvec{v}$ defines the velocity of the Gaussian wave packet, $A_0$
the strength of the interaction, and the parameter $\tilde t$ is needed
to define and calculate the derivatives for energy and momentum in
(\ref{theory:fieldParticle:fieldEnergy}) and
(\ref{theory:fieldParticle:fieldMomentum}).  The three position
arguments $x_i$ are fixed by the interaction position.

\begin{figure}
\centering
\includegraphics[width=0.7\textwidth,keepaspectratio]{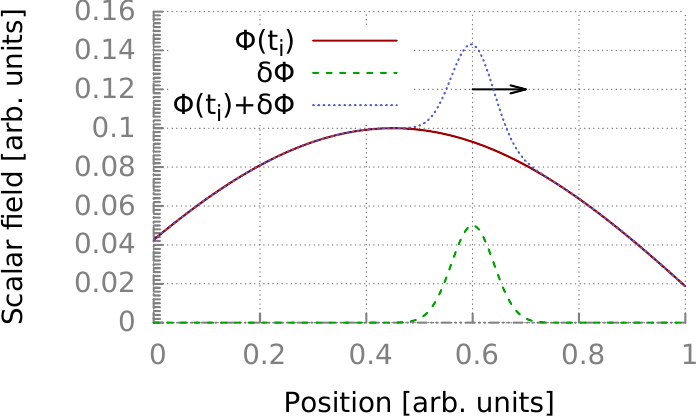}
\caption{(Color online) Visualization of the interaction principle, as described in
  (\ref{eq:parameterization:gauss3D}). The energy of the initial field
  $\phi(t)$ is changed by a parameterized field variation $\delta
  \phi$. The resulting field $\phi(t) + \delta \phi$ is increased by a
  given energy $\Delta E$ and momentum $\Delta {\bvec P}$. The traveling direction of the Gaussian is depending on the momentum.
  \label{fig:equilibriumBox:fieldPlusInteraction} }
\end{figure}

To find $A_0$ and $\bvec{v}$ in the parameterization
(\ref{eq:parameterization:gauss3D}), which solve
(\ref{eq:theory:FieldManipulationEnergy}) and
(\ref{eq:theory:FieldManipulationMomentum}) for a given $\Delta E$ and
$\Delta \bvec{P}$, the four coupled and non-linear equations have to be
solved with a numerical equation solver with $\Delta E$ and
$\Delta \bvec{P}$ given by the physical interaction; the definition of
these quantities will be given in various examples in the following
Sections. A simple visualization of this principle is given in Figure
\ref{fig:equilibriumBox:fieldPlusInteraction}, which shows the local
modification of the field by a Gaussian.

The three widths of the Gaussian, $\sigma_i$, are free parameters and
can be fixed to a single spherical radius by
$\sigma_x = \sigma_y = \sigma_z = \sigma$. It determines the
interaction volume and should be chosen to fit the system scale. It has
an impact on the minimal scale of possible modes in the system, as we
will see in (\ref{eq:cuttofSigma}).

Furthermore, the parameterization (\ref{eq:parameterization:gauss3D})
should be seen as a non-relativistic approximation which is valid for
small velocities.  The parameterization is not Lorentz invariant,
resulting in spatial extent which does not depend on the velocity. This
has the interesting effect, that for a (1+3)D system the maximal
momentum to energy ratio which can be generated with
(\ref{eq:parameterization:gauss3D}) is
$\max \left \{ \frac{P}{E} \right \} = 1/2$. The parameterization can
be extended with a Lorentz boost, for example along the $x$-direction
for $\bvec{v} = \left(v_x, 0, 0 \right)$
 \begin{equation}
\begin{split}
\label{app:3DgaussiannRel}
\delta \phi(\bvec{x}, t) = &A_0  \exp \left(-\frac{\gamma^2\left( x
      -v_x t \right)^2}{2 \sigma^2}\right) \\
&\times \exp \left(-\frac{y^2 + z^2}{2 \sigma^2}\right)
\end{split}
 \end{equation}
 The Lorentz boost leads to a disc-shaped deformation of the initially
 spherical Gaussian. With the boost, the momentum to energy ratio of
 (\ref{app:3DgaussiannRel}) has the correct relativistic limit.
\begin{equation}
\lim_{v \to 1} \left \{ \frac{P}{E} \right \} = 1
\end{equation}
At $v=0.3$ both solutions differ by a factor of about $18\%$, for small
velocities they are nearly the same, and
(\ref{eq:parameterization:gauss3D}) can be used as a safe approximation.

\subsection{Modeling of energy transfers}

In Sect.\ \ref{sec:EnergyTransfer}, the mathematical framework for
energy and momentum transfer has been discussed.  To apply this method
to physical systems, additional modeling has to be done. The above
method describes, how to transfer a given amount of energy $\Delta E$ and
momentum $\Delta \bvec{P}$ to or from a scalar field at a given
interaction point $\bvec{x}$ and time $t_i$. To realize this method in a
transport or Monte-Carlo simulation, appropriate values for $\Delta E$,
$\Delta \bvec{P}$, $\bvec{x}$ and $t_i$ have to be defined.  Motivated
by the stochastic interpretation of interaction cross sections, a
probability distribution for energy and momentum exchanges,
\begin{equation}\label{theory:interactionProbability}
 P \left ( \Delta E, \Delta \bvec{P}, \Delta t \right),
\end{equation}
can be defined, which has to be derived according to the physical model;
in Sect.\ \ref{sec:TestCalculations} we will give various examples,
calculations and results for such a modeling. In general, the
probability distribution (\ref{theory:interactionProbability}) can have
a memory kernel and can depend on the whole history with
$P \left ( \Delta E, \Delta \bvec{P}, t \right)$, but we will use the
Markov approximation in which only the current system state is important
for future events, and (\ref{theory:interactionProbability}) depends on
the time-step size, $\Delta t$, only.

\section{\label{sec:TestCalculations}Examples and Model Calculations}

\subsection{\label{sec:DampedHarmonicOssci}Discretely damped harmonic oscillator}

As the most simple test system, we choose the classical one-dimensional
oscillator with damping,
\begin{equation}
\label{harmoincOscillator:EOM1}
\ddot x(t) + \gamma \dot x(t) + \omega_0^2 x = 0.
\end{equation} 
In this example the mass of the system is set to unity. With no loss of
generality units like $c$, $\hbar$ and $k_B$ are set to unity in the
rest of the paper, and physical quantities like the energy are chosen to
have an arbitrary unit (arb. units).

The frictional part $\gamma \dot{x}$ dissipates energy from the system
in a continuous process. We want to model the same system to have a
discrete, noncontinuous damping such that, within an ensemble average,
both systems should behave the same.  We assume to have a weak damping
of the oscillator with $\gamma < \omega_0$.  The analytic solution of
the oscillator's energy scales with
\begin{equation}
  E(t) \approx \bar E_0 \ \mathrm{e}^{-\gamma t}.
\end{equation} 
So the linear damping term $\dot x$ leads to an exponential energy
loss in the oscillator. To model a discrete damping for
(\ref{harmoincOscillator:EOM1}), we could introduce a deterministic
formalism which removes a given quantum of energy at fixed times out of
the system. The second and more natural choice is a probabilistic
ansatz, which models the system's initial total energy $E_0$ as a sum of
small energy quanta $\overline{\Delta E}$,
\begin{equation}
  E_0 = N_0 \cdot \overline{\Delta E},
\end{equation}
where $N_0$ is the initial number of energy quanta and can also be
called ``steppiness'' because it defines in which energy steps the
system can be damped.  The energy of the system can be damped by
changing the number of energy quanta $N(t)$
\begin{equation}\label{eq:energyQuantization}
  E(t) = N(t) \cdot \overline{\Delta E}
\end{equation}
This change of $N(t)$ has to be modeled according to the equations of
motion.  We now explain how to find an interaction-probability function
like (\ref{theory:interactionProbability}) for this system. In this
example, we can assume a two-state interaction: for a given $\Delta t$,
the oscillator can lose a quantum of energy $\overline{\Delta E}$, or it
can be left undisturbed. For $\Delta t \ll 1$ we can neglect multiple
decays; additionally we assume a Markov process, so the oscillator only
depends on its current state and has no ``memory''.  Using these
constraints, the interaction-probability distribution
$P\left ( \Delta E, \Delta t \right)$ without memory kernel can be
described as
\begin{equation}
  P\left ( \Delta E, \Delta t \right) = \Pr_{\textrm{loss}}(\Delta t)\delta(\Delta E - \overline{\Delta E}) + \Pr_0(\Delta t)\delta(\Delta E )
\end{equation}
with $\Pr_{\textrm{loss}}$ being the probability to lose an energy
quantum ${\Delta E}$ in the time interval $\Delta t$, while $\Pr_0$ is
the probability for the system to stay unchanged. Both probabilities are
related by the normalization of the probability distribution,
\begin{equation}
\int P\left ( \Delta E, \Delta t \right) \dd \Delta E = \Pr_{\textrm{loss}}(\Delta t) + \Pr_0(\Delta t) = 1.
\end{equation}
To find the probability for the oscillator to lose a certain amount of energy, we assume
that every energy quantum $\overline{\Delta E}$ can decay independently.
The definition for the exponential decay is
\begin{equation}
  \frac{\dd N(t)}{\dd t} = -\gamma N(t)
\end{equation}
with each decay event having a constant and independent decay probability in a time step $\dd t$ of $\Pr = \gamma \dd t$.
With $\dd t \to \Delta t$ and $\Delta t \ll 1$ we can write
\begin{equation}
  \Delta N(t) = - \gamma \Delta t N(t) 
\end{equation}
However, we want to calculate the probability of a single energy quantum to decay.
The number of energy quanta is given by (\ref{eq:energyQuantization})
\begin{equation}
  N(t) = E(t) / \overline{\Delta E} 
\end{equation}
which increases the number of energy quanta if $\overline{\Delta E}$ decreased. 

The total probability of a decay of a single quantum in a system of many
quanta in first-order approximation factors to the product of an
individual quantum's decay probability times the number of individual
quanta,
\begin{equation}\label{PropabilityScalarDamping}
	\Pr_{\textrm{loss}} \left ( \Delta E \right) =
   \gamma \cdot \Delta t \cdot N(t) = \gamma \cdot \Delta t \ \frac{E(t)}{\overline{\Delta E}} \ .
\end{equation}
For $N_0 \to \infty$ or $\Delta E \to 0$ this is the definition of the
exponential decay law, while a finite $N_0$ will give a discrete
exponential decay for a finite ensemble. For the total
probability-distribution function we obtain
\begin{equation}
\begin{aligned}
  P\left ( \Delta E, \Delta t \right) = & \ \delta(\Delta E - \overline{\Delta E}) \left(\gamma \cdot \Delta t \frac{E(t)}{\overline{\Delta E}} \right) \\
  & + \delta(\Delta E ) \left( 1 -\gamma \cdot \Delta t \frac{E(t)}{\overline{\Delta E}} \right) \ .
\end{aligned}
\end{equation}
Simulating $P\left ( \Delta E, \Delta t\right)$ will give the same
average energy loss scaling for $E(t)$ as the original harmonic
oscillator with continuous damping.

In a numerical realization, the oscillator is propagated with the free
equation of motion,
\begin{equation}
\frac{\dd^2 x}{\dd t^2} + \omega_0^2 x = 0.
\end{equation}
This equation of motion conserves the total energy. To simulate damping,
at every time step the decay-probability density
$P\left ( \Delta E \right)$ is sampled using Monte-Carlo techniques. In
case of a decay, the oscillator will lose the given amount $\Delta E$ by
employing the method described in section \ref{sec:EnergyTransfer}. In
case of an oscillator, only the energy equation
(\ref{eq:theory:FieldManipulationEnergy}) has to be solved. The change
$\delta x$ on $x(t)$ becomes a simple shift of the oscillator,
 \begin{equation}
  x_t \to x_t + \delta x.
 \end{equation}
 For a harmonic potential, this can be done analytically by solving
\begin{equation}
\begin{split}
\label{eq:deltaDiffEnergy}
\overline{\Delta E} &= E(x_{t+1}) - E(x_t) \\
&= \frac{1}{2}\left [ \omega_0^2 x_{t+1}^2 + \dot x_{t+1}^2 - \omega_0^2
  x_{t}^2 - \dot x_{t}^2 \right].
\end{split}
\end{equation}
The derivatives are approximated by the first-order difference
discretization and with $\dd t \to \Delta t$:
\begin{equation}
    \dot x_{t+1}  = \frac{x_{t+1} - x_t}{\Delta t}
\end{equation}
Solving (\ref{eq:deltaDiffEnergy})for $x_{t+1}$ results in a rather
lengthy equation which can be simplified by neglecting all terms of
order $\mathcal{O} \left( \omega^2 \Delta t^2 \right)$ and higher. The
result is
\begin{equation}\label{eq:changeEnergyScalar}
  x_{t+1} = x_{t} \pm \Delta t \sqrt{2 \overline{\Delta E} + \dot x_t^2}.
\end{equation}
This can be seen as an addition to the undisturbed equations of
motion. With $\overline{\Delta E} \to 0$ equation
(\ref{eq:changeEnergyScalar}) becomes the usual, first order Euler
propagation for a differential equation:
$x_{t+1} = x_t + \Delta t \cdot \dot x_t$. The additional term
$\overline{\Delta E}$ is the change of the system given by the
``interaction kick'', changing the total system's energy with exactly
this amount of energy. The sign $\pm$ in front of the square root is
determined by the direction of $\dot x_t$, a kick with
$\overline{\Delta E}>0$ should always point in the direction of the
current velocity $\dot x_t$. Note that $\overline{\Delta E}$ can always
be positive while it can only be negative if
$\dot x_t^2 - 2 \left |\overline{\Delta E} \right | > 0$ to have a real
solution for the propagation equation.

\begin{figure}
    \centering
    \includegraphics[width=0.7\textwidth]{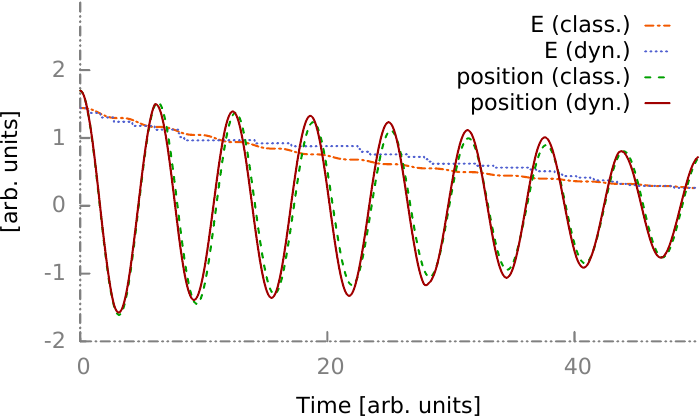}
    \caption{(Color online) Simulation of a discretely damped harmonic
      oscillator. The number of simulated particles and energy
      excitations is $N=150$. With every decay of a particle, the
      oscillator loses a bit of energy, which leads to damping of its
      motion. The discrete and continuous version have the same ensemble
      average. In a single run statistical fluctuations occur.}
 \label{fig:appendix:expoentialDecay:DifferentParticlesScalar}
\end{figure}

Fig.\ \ref{fig:appendix:expoentialDecay:DifferentParticlesScalar} shows
a single simulation run of such a system.  Both curves show statistical
fluctuations but are in good agreement over the runtime. Averaged over
many realizations, the energy shows the expected exponential
decay. Using different values for $N$, the `steppiness' can be
controlled from a very smooth damping to a rough damping with only a few
steps. On average, all show the exponential decay with the expected
rate.

\subsection{Harmonic oscillator coupled to a heat bath (Langevin approach)}

In a second benchmark, we extend the damped harmonic oscillator by a
coupling to a heat bath via a Langevin equation, similar to
\cite{feynman1963theory}. This calculation is done for a simple harmonic
oscillator as well as for a scalar field in (1+1) space-time dimensions.
Again, the mass of the system is set to unity, and energy units labeled
as arbitrary unit (arb. units).  For the non-relativistic case, the equation
of motion reads
 \begin{equation}\label{eq:EOM:langevinosci}
  \frac{\dd^2}{\dd t^2} x(t) + \gamma
  \frac{\dd}{\dd t} x(t) + x(t) = \kappa \xi(t),
\end{equation}
where $\xi(t)$ is defined as Gaussian white noise with
$\langle \xi(t) \xi(t') \rangle = \delta(t-t')$, using the equipartition
theorem and the fluctuation-dissipation theorem
\cite{toda1991statistical}, we can fix the strength of the stochastic
force in the equilibrium case as
\begin{equation}\label{eq:dissipationFluctuationOscillator}
 \kappa = \sqrt{{2 \gamma T}}.
\end{equation}
This equation of motion describes a damped harmonic oscillator, driven
by a Gaussian-distributed random force $\xi(t)$, which can increase or
decrease the energy of the system by ``kicking'' the oscillator. On
average, the oscillator will show a Gaussian position distribution, and
by using the Fokker-Planck equation one can derive the equilibrium
distribution for the energy, which is a Boltzmann distribution
\begin{equation} f(E) \propto \exp(-E/T). \end{equation}
To simulate the oscillator with the Langevin-dynamics within our proposed method,
we propagate the system with the interaction-free equation of motion,
\begin{equation}
  \frac{\dd^2}{\dd t^2} \ x(t) + x(t) = 0,
\end{equation}
and model the discrete interactions again as small kicks to have the
same statistical averages as the original Langevin equation of motion
(\ref{eq:EOM:langevinosci}). A representation for the stochastic force
$\kappa \cdot \xi(t)$ in terms of energy changes can be found by using
the energy-work theorem, applied to the fluctuating part of the force in
the Langevin equation,
\begin{equation}
 \frac{\dd E}{\dd t} = \dot x(t) \cdot F(t) = \dot x(t) \cdot \kappa \cdot \xi(t),
\end{equation}
which after discretization with $\dd t \to \Delta t$ and $t \to t_n$
becomes
\begin{equation}\label{deltaE.langevin}
 \Delta E = \dot x(t) \cdot \Delta t \cdot \kappa \cdot \tilde \xi(t_n).
\end{equation}
The stochastic force $\tilde \xi(t)$ is still a Gaussian white noise
with a normally distributed random number $\xi_n$ at each time step
$t_n$,
\begin{equation}
  \tilde \xi(t_n) = \frac{\xi_n}{\sqrt{\Delta t}}.
\end{equation}
The factor ${\Delta t}^{-\frac{1}{2}}$ is needed to fix the norm of the
uncorrelated white noise via
\begin{equation}
  \langle \xi_n \xi_m \rangle = \frac{\delta_{mn}}{\Delta t}.
\end{equation}
The total interaction-probability distribution function,
$P\left (\Delta E, \Delta t \right)$, is now composed of four components
of single probabilities.  The probability of no interaction in a time
interval $\Pr_0$, the damping of the oscillator by the process
$\gamma \dot x(t)$ with the probability $\Pr_{\textrm{loss}}$, as
discussed in the previous section, and the two cases where the energy of
the system is changed by the stochastic force $\tilde \xi(t)$: we can
see in (\ref{deltaE.langevin}) that depending on the signs of $\xi$ and
$\dot x(t)$ a random kick can add energy to a system or dissipate energy
from it. Both processes are symmetric in general and from this symmetry
one obtains $\langle \Delta E \rangle=0$ in equilibrium. However
$\langle \Delta E^2 \rangle > 0$ is always given.

The probability $\Pr_{\textrm{loss}}$ has already been discussed in the
previous section \ref{sec:DampedHarmonicOssci} and follows the same
schematics here. Due to their symmetry, the loss and gain terms induced
by the stochastic force, given by (\ref{deltaE.langevin}), can be
subsumed in a single probability density term. The sum of all terms for
the interaction probability distribution is
\begin{equation}\label{eq:sum:ProbDistLangevin}
 \begin{aligned}
   P(\Delta E, \Delta t) = &\delta \left(\Delta E - \dot x(t) \cdot \Delta t \cdot \kappa \cdot \xi(t) \right) \\
   & + \delta(\Delta E - \overline{\Delta E}) \left(\frac{\gamma \cdot \Delta t}{\overline{\Delta E}} E(t) \right) + \Pr_{0} \ \delta(\Delta E).
  \end{aligned}
\end{equation}
The no-interaction probability $\Pr_0$ is again fixed by the normalization
condition, \linebreak $\int P(\Delta E, \Delta t) \dd \Delta E =1$. Equation
(\ref{eq:sum:ProbDistLangevin}) looks quite complicated, but the single
terms can be easily interpreted and simulated. The last term is the
contribution for no interaction to happen at all. The second last term
describes the probability for a system to lose a given amount of energy
$\overline{\Delta E}$ due to friction, just like in our first example.
Finally, the first term describes the process of changing the energy
from a kick by the stochastic force. The system gains energy, if the
direction of the kick points in direction if the velocity,
$\dot x(t) \cdot \xi(t) > 0$, and loses energy if the kick reduces the
velocity, which is the case for $\dot x(t) \cdot \xi(t) < 0$.

To sample the gain and loss terms by the random kicks in the probability
density, one can simply sample a random kick $\xi_n$ and calculate the
given energy difference from (\ref{deltaE.langevin}), which is
propagated back to the system. As we simulate a simple (0+1)D problem,
the already known relation for changing the energy
(\ref{eq:changeEnergyScalar}) can be used.

In this example we describe the dissipative process with discrete decay
steps while the energy fluctuations given by $\xi(t)$ can have
continuous values. Even though this seems contradictory, it has two
reasons: we wanted to stay as close as possible to the original Langevin
equation, which has continuous interactions. The second reason is that
we wanted to introduce continuous values for $\Delta E$ at this point
because in the last example in this paper
(\ref{seq:ParticleFieldMethod}) particles and fields exchange energy by
discrete particle annihilation and creation processes. While these
processes are discrete in time, their energy and momentum spectrum is
continuous.

\begin{figure}
\centering
\includegraphics[width=0.7\textwidth]{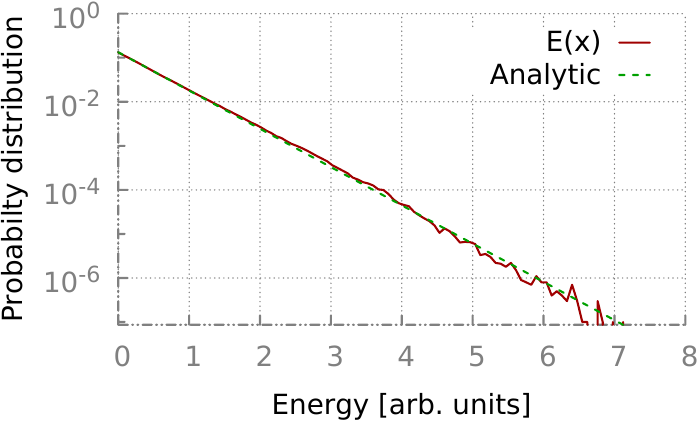}

\includegraphics[width=0.7\textwidth]{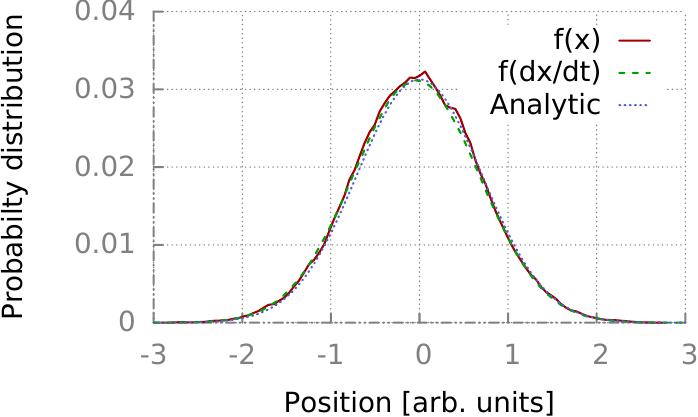}

\caption{(Color online) Simulation of a simple harmonic oscillator,
coupled to a heat bath via a Langevin equation. The random force and the
damping is implemented with the noncontinuous energy-transfer method.
The plot shows the distribution function of the energy, the position
distribution and the velocity distribution for a temperature of $T=0.5
\textrm{arb. units}$ and $10^8$ calculation steps and a corresponding
energy-step size. \tabularnewline Both plots coincide with the expected
analytical result, which is $\exp\left(-E/T\right)$ for the energy
distribution (top) and $1/\left(\sqrt{2 \pi T} \right) \exp\left( -m
\omega_0^2 x^2/2T \right)$ for the spatial distribution.
\label{fig:appendix:expoentialDecay:DifferentParticles}
}
\end{figure}
\begin{figure}
  \centering \centering
  \includegraphics[width=0.7\textwidth]{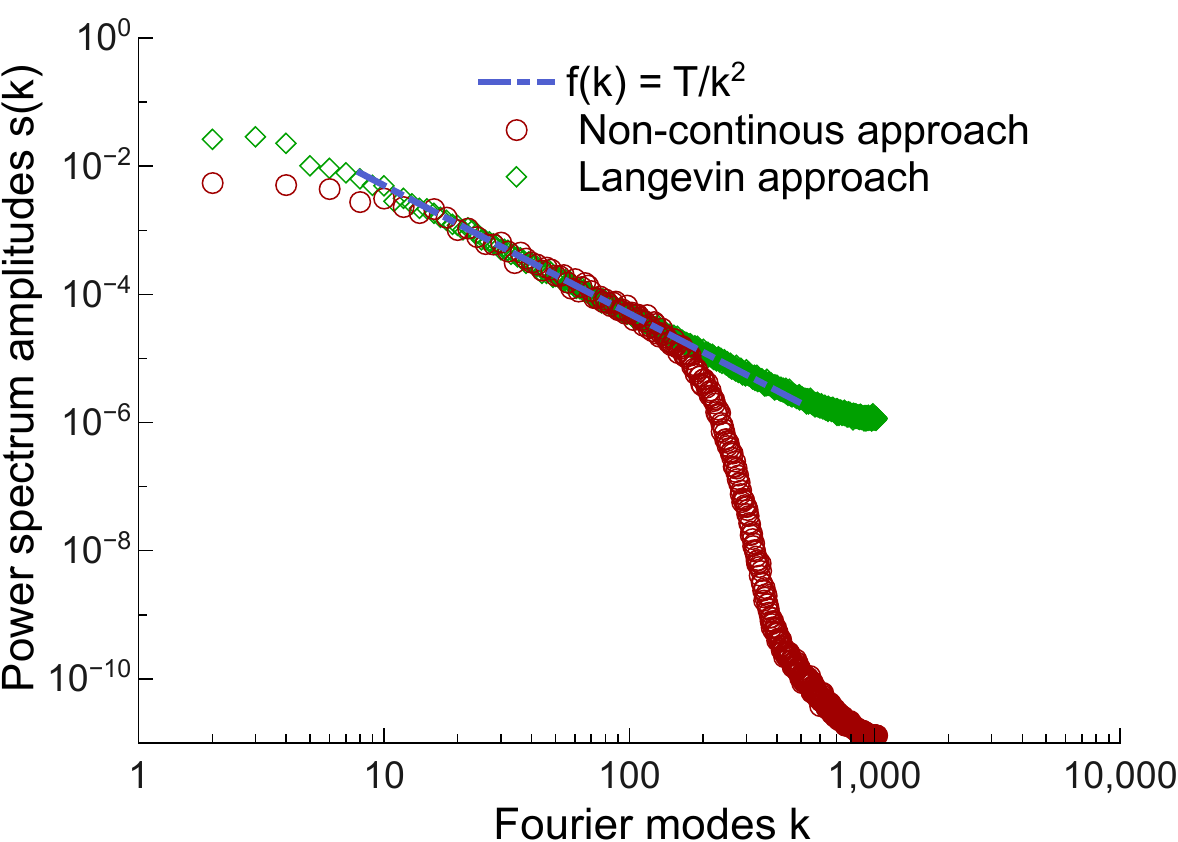}
  \caption{(Color online) Time averaged power-spectrum of the scalar
field $\phi(x, t)$ in (1+1) space-time dimensions.  The spectrum is
calculated according to (\ref{eq:DefinitionPowerSpectrum}).  The system
is a one-dimensional harmonic oscillator, coupled to a Langevin equation
with a continuous and a noncontinuous ansatz. The dashed, blue line
shows the theoretical solution of the system $T/k^2$, the red circles
the noncontinuous approach and the green diamonds show the classical
Langevin equation as a reference. All three curves are in very good
agreement in the intermediate region and show the same system
temperature, which was chosen to $T=0.5 \textrm{arb. units}$ Both Fourier
plots are averaged over 100 snapshots of the same run.  Simulated was a
grid with 1024 points.  At large wavelengths, both systems deviate due
to finite system size effects. At small wavelengths the noncontinuous
methods has an interesting cut-off due to the finite-size of the
interaction region in comparison to the point-like interactions of the
classical Langevin method.
  \label{fig:1DLangevin:FullFigure}
}
\end{figure}
Fig.\ \ref{fig:appendix:expoentialDecay:DifferentParticles} shows the
position and velocity distribution of the harmonic oscillator, which is
a Gaussian.  The width of the distribution depends on the temperature in
the fluctuation-dissipation relation
(\ref{eq:dissipationFluctuationOscillator}) and scales with $\sqrt{T}$.
The energy distribution shows an exponential tail $\exp(-E/T)$ with the
same temperature. The numerical calculation was performed with our
method, which does not solve exactly the classical Langevin equation,
but a set of equations which have the same statistical description as
explained above. As a result, the statistical properties of the
simulated system are in good agreement due to the consistent description
of the energy-exchange rates.

\subsection{One-dimensional scalar field coupled to Langevin equation}

The previous calculations in this work have dealt with a simple harmonic
oscillator. In the next example, the scalar system
(\ref{harmoincOscillator:EOM1}) is advanced to a $1+1$-dimensional case
with a spatially extended field. The position coordinate of the
oscillator, $x(t)$, is now generalized to a scalar field $\phi(x,t)$.
The equation of motion of a potential-free Klein-Gordon equation with a
stochastic force and damping is given by
\begin{equation}
\begin{split}
\label{eom:Langevin1D}
 \left( \frac{\partial^2}{\partial t^2} - \frac{\partial^2}{\partial
    x^2} \right) \phi(x, t) + \gamma \frac{\partial}{\partial t} \phi(x, t) + \phi(x, t) = \kappa
\xi(x, t)
\end{split}
\end{equation}
The stochastic force is the $1+1$-dimensional extension of the Gaussian
white noise \linebreak
$\langle \xi(x,t) \xi(x', t')\rangle = \delta(x-x') \delta(t-t')$. The
stochastic process (\ref{eom:Langevin1D}) is simulated as a reference
system. In comparison, in our simulation the field $\phi(x,t)$ is
propagated with the disturbance-free equations of motion. The stochastic
force and damping are simulated analogously as in the (0+1)-dimensional
example. Because of the additional spatial dimension of the system, we
could model a momentum exchange $\Delta P$, as given in
(\ref{theory:interactionProbability}). We neglect this term because the
original Langevin-equation does not have this term either. However, the
last example in this paper will address this issue.

In the previous example of the simple oscillator, the change of the
oscillator's position $\Delta x$ could be calculated with the analytic
relation (\ref{eq:changeEnergyScalar}), which included both the kick
from the energy exchange and the propagation by the equation of
motion. Such an analytic relation can not be found anymore in this
example because the spatial extent of the system and the interaction
region leads to a complex dependency of the energy equation
(\ref{eq:theory:FieldManipulationEnergy}), requiring a numerical
solution of the problem. The numerical problem decomposes in finding a
solution for the kick by the change of energy and then for propagating
the newly found field-configuration with the equation of motion.
Additionally, the change of the field $\Delta \phi(\bvec x)$ has to be
parameterized, as many possible solutions could be found for
eq. (\ref{eq:theory:FieldManipulationEnergy}) for systems with a spatial
extend. In this example a one-dimensional form of the Gaussian
parameterization (\ref{eq:parameterization:gauss3D}) has been chosen.

For each point of the system an interaction probability is sampled with
(\ref{eq:sum:ProbDistLangevin}). In case of an interaction, the center
of the Gaussian interaction parameterization is located at the
interaction point, and the change of the field
$\phi(\bvec{x}, t) + \delta\phi(\bvec{x})$ is solved by using equation
(\ref{eq:theory:FieldManipulationEnergy}) with help of a numerical
solver. For every interaction at some point $\bvec x$ the neighbor cells
of the interaction point are changed as well. The reason is the spatial
extension of the Gaussian parameterization leading to a smeared
interaction region or volume. This holds both for adding energy as well
as removing energy from the system.

As shown in figure \ref{fig:equilibriumBox:fieldPlusInteraction}, this
will result in a field given by a superposition of small interaction
``bumps''. In the equilibrium state, the field distribution will show
spatial fluctuations, distributed among the field's modes. Overall the
equilibrium state behaves very similar to the oscillator case but with a
spatial extent in one dimension and therefore more dynamics. The same
holds true for the interaction probability-distribution density, which is
the spatially extended version of (\ref{eq:sum:ProbDistLangevin}):
\begin{equation}\label{eq:sum:ProbDistLangevin1D}
\begin{aligned}
    P(x, \Delta E, \Delta t) =    \delta \left(\Delta E - \dot \phi(x, t) \cdot \Delta t \cdot \kappa \cdot \xi(x, t) \right) \\
   + \delta(\Delta E - \overline{\Delta E}) \left(\frac{\gamma \cdot \Delta t}{\overline \Delta E} E(x, t) \right)  + \Pr_{0} \ \delta(\Delta E).
\end{aligned}
\end{equation}
A very interesting observable is the power spectrum of the field,
\begin{equation}\label{eq:DefinitionPowerSpectrum}
  S(\bvec k) = \lim_{t \to \infty} \frac{1}{2 t} \int_{-t}^{t} 
  \dd t \left | \mathcal{F}[\phi(x,t)]({\bvec k}) \right |^2,
\end{equation}
with the spatial Fourier transformation $\mathcal{F}$
\begin{equation}
  \mathcal{F}[\phi](\bvec k, t) = \int \dd^3 \bvec x \ \exp \left( - i \bvec k \cdot \bvec x \right) \phi(\bvec x, t) \ .
\end{equation}
In case of the classical Langevin equation, the expected distribution of
the power spectrum can be calculated. A damped field coupled to a
white-noise process is expected to show Brownian noise, as it
effectively integrates the white noise over time
\cite{lindstrom1993fractional}. The resulting field has an average power
spectrum with the temperature as the mode amplitude
\cite{wienerTheorem,toda1991statistical},
\begin{equation}
\label{brownian-power-spec}
 S(k) = \frac{T}{k^2}.
\end{equation}

Fig.\ \ref{fig:1DLangevin:FullFigure} shows the expected spatial
spectrum for the classical Langevin equation (\ref{eom:Langevin1D}) and
for the simulation with our proposed method. For large and small momenta
$k$, deviations occur due to finite-system effects, in the intermediate
region the spectrum shows a very good agreement with
(\ref{brownian-power-spec}).  At some point the small wavelengths with
large $k$ are suddenly strongly suppressed.

The explanation for this behavior is the fact that we use a finite
volume excitation in the parameterization
(\ref{eq:parameterization:gauss3D}).  In a classical Langevin equation,
the point-like stochastic force $\xi$ has a constant spectrum
$S_\xi(k) \sim c$, allowing to excite any modes.  Within our method, the
energy is changed in a small but finite-size sub-volume. The smallest
excitable mode in the system has therefore the same scale as the
interaction volume. Within the Gaussian parameterization
(\ref{eq:parameterization:gauss3D}), this scale is the width
$\sigma$. The resulting mode cutoff can be calculated evaluating the
spectrum of the parameterization (\ref{eq:parameterization:gauss3D}),
here for our (1+1)-dimensional example:
\begin{equation}\label{eq:cuttofSigma2}
 \left | \mathcal{F}[\delta \phi(x,t)](k) \right |^2 \sim e^{- k^2 \cdot \sigma^2 / 2} \approx 
 \begin{cases}
     1 & \text{for } k \ll \sigma, \\
     0 & \text{for } k \gg \sigma.
   \end{cases}
\end{equation}
At the scale $k^2 \sigma^2 \approx 1$ the Gaussian shape leads to a soft
cutoff which suppresses all higher modes, preventing the well-known UV
catastrophe of classical thermal fields. We therefore define a soft
cutoff scale at
\begin{equation}\label{eq:cuttofSigma}
  k_{\mathrm{Cutoff}} \equiv \frac{\sqrt{2}}{\sigma} \ .
\end{equation}

These results show that our method is capable of simulating a thermal
system with Langevin dynamics with controlled systematic numerical errors.

\subsection{Particle ensemble coupled to scalar field}
\label{seq:ParticleFieldMethod}

In this Section we use the proposed method to couple an ensemble of
particles to a scalar field in a (3+1)-dimensional simulation. The
simulated system is a microcanonical box with a scalar field, particles
and anti-particles. Particles can perform two-body elastic collisions,
the field is propagated via a wave equation, like in the sections
before. Additionally, field and particles can interact by microscopic
particle annihilation and creation processes. This system is motivated
by a physical problem which is discussed in an upcoming paper. However,
we want to use these calculations to demonstrate the proposed method in
a more complex example and try to keep the details of the used model as
generic as possible, while still being exact in our derivations. The
underlying physical system is the linear sigma model with constituent
quarks. We model quark-pair creation and annihilation to study
non-equilibrium effects near and at the phase transition. The process of
creating particles from field modes and vice versa is crucial for
describing critical phenomena and fluctuations driven by the dynamics of
the phase transition.

The equations of motion in a general form are
\begin{equation}
\begin{split}
  \left ( \frac{\partial^2}{\partial t^2} - \nabla^2_{{\bvec x}} \right)
  \phi({\bvec x}, t)
  = &\lambda_1 \phi^3({\bvec x}, t) + \lambda_2 \phi({\bvec x}, t) \\
  &+ U_{\bar q q}({\bvec x}, t) + U_0
\end{split}
\end{equation}
with the coupling strength for the potentials $\lambda_1, \lambda_2
\gg 1$, a source term $U_0$ and a mean field potential between the field
and the particles $U_{\bar q q}$. Besides the mean-field and potential
interactions, a particle $q$ and an anti-particle $\bar q$ can
annihilate to a field-quantum. The underlying process is given by the Yukawa
coupling,
\begin{equation}
  \mathcal{L}_\textrm{int} = g \bar \sigma \phi^* \phi,
\end{equation}
which can create a $\sigma$ particle by pair annihilation,
$\bar q + q \to \sigma$, or destroy it in the inverse decay process,
$\sigma \to \bar q + q $.

The interaction between particles and the field is modeled in several
steps. The quark-annihilation process is calculated with a microscopic
cross section, the generated particle is treated as an unstable particle
resonance, the $\sigma$ particle. The created $\sigma$ particle is not
propagated or added to the system as a real particle, but its energy and
momentum are transferred to the scalar field, keeping the system's total
energy and momentum conserved.  In the inverse process, excitations in
the scalar field are treated as energy excitations. These excitations
are modeled as unstable particles which can decay to a quark-antiquark
pair.


The interaction probability of the two incoming particles for the
process $\bar q q \to \sigma$ is calculated microscopically. In our
simulation it is modeled with a constant, isotropic cross section
$\hat \sigma_{\bar q q \to \sigma}$ with respect to all kinematic
constraints. For Monte-Carlo sampling, we use the stochastic
interpretation of the cross section \cite{Xu:2004mz} for a set of
particles in a cell employing a constant and isotropic cross section:
\begin{equation}\label{eq:collisionProbability}
 \Pr\left( \bar q q \to \sigma \right) = \hat \sigma_{\bar q q \to \sigma} \ v_{\textrm{\textrm{rel}}} \frac{\Delta t}{\Delta V N_{\textrm{test}}}
\end{equation}
with the particles' relative velocity,
\begin{equation}
v_{\textrm{\textrm{rel}}} =   \frac{s}{2 E_1 E_2},
\end{equation}
and the Mandelstam variable,
\begin{equation}
s = \left( p^1_\mu + p^2_\mu \right) ^2.
\end{equation}
We have chosen the Breit-Wigner cross section \cite{PhysRev.49.519,
  PhysRevD.71.085018}
\begin{equation}\label{eq:BreitWignerCrossSection}
\sigma_{\bar q q \to \sigma}(s) =  \frac{\bar \sigma \ \Gamma}{\left( \sqrt{s} - m_\sigma \right)^2 + \left( \frac{1}{2} \Gamma \right)^2 }
\end{equation}
with a constant factor $\bar \sigma$ and the mass of the created unstable particle $m_\sigma$.

given by
\begin{equation}
\Delta E = \sqrt{s}, \quad \Delta \bvec P = \bvec{p}_1 + \bvec{p}_2 \,.
\end{equation}
At the particles' interaction point the particles are removed from the
ensemble and their total energy and momentum are transferred to the
field $\phi$ at this point, keeping the total energy and the total
momentum conserved. This transfer is done by changing the field energy
and momentum at the interaction point of the particles using our
proposed method. The energy and momentum difference equations
(\ref{eq:theory:FieldManipulationEnergy}) and
(\ref{eq:theory:FieldManipulationMomentum}) are solved for the
interaction-time step with a numerical solver. The field
$\phi({\bvec x}, t)$ is changed by employing the 3D Gaussian
parameterization $\delta \phi({\bvec x})$
(\ref{eq:parameterization:gauss3D}). By changing the amplitude as well
as the direction of motion of the Gaussian, both the energy and momentum
can be changed within the interaction volume until $\Delta E$ and
$\Delta \bvec P$ are transferred to the field. Figure
\ref{fig:equilibriumBox:fieldPlusInteraction} shows a simplified version
of this process in which small Gaussian blobs over a small volume
generate small energy excitations on the field $\phi$.

To guarantee thermal and chemical equilibration, the inverse process has
to be implemented according to the principle of detailed balance. This
has several implications. First, the average interaction rates of
particle creation and annihilation has to be the same for a given
temperature, leading to no change of net-particle number. Secondly, the
average energy exchange per process has to be the same for both
processes. Finally, the spectra of both processes have to be the same.

We have already discussed the method for particle annihilation. The
inverse process, particle production, has to be modeled differently,
because the field has no initial particles which we could use for
Monte-Carlo sampling of a collision term. Instead, we only have the
scalar field $\phi$ and its properties like energy and momentum density,
from which we have to derive particle-like properties. This step is
again subject to the underlying physical model. For every point in
space, we assume the field excitations to consist of unstable particles
which can decay to stable particles, $\sigma \to \bar q q$. In case of a
decay, the field $\phi$ loses the amount of energy at the interaction
point, leading to an effective damping of the field. This decay process
is modeled in two steps. First we have to assume a distribution function
$f_\sigma(\bvec x, \bvec p, t)$ for $\sigma$ particles at every possible
interaction point. The properties of $f_\sigma(\bvec x, \bvec p, t)$
have to be derived from the field properties at every point in space
which is done by assuming local equilibrium via coarse graining within a
field cell. To be consistent with detailed balance, the equilibrium
distribution for $f_\sigma$ must have the same temperature as the
particle's distributions, $f_q$ and $f_{\bar q}$. In the linear $\sigma$
model, the potential and therefore the equilibrium-mean field has a
thermodynamical temperature dependence, which maps a mean-field value
for every temperature $T \to \langle \phi \rangle(T)$. By inverting this
relation $\langle \phi \rangle(T) \to T$, we can calculate the effective
temperature of the field at every point of the field. The resulting
phase-space distribution function, $f_{\sigma}$ defines the particle
density by
\begin{equation}
  n_\sigma\left( \bvec x, t \right) = \int \frac{\dd^3 \bvec
    p}{(2\pi)^3} f_\sigma \left (\bvec x, \bvec p, t \right).
\end{equation}
Depending on the underlying distribution function $f_\sigma$ a relation
between the particle and energy density has to be found
\begin{equation}
  n_\sigma \left( \bvec x, t \right) \to n_\sigma \left( \bvec x, t, \epsilon \left( \bvec x \right) \right) .
\end{equation}
The energy density $\epsilon$ is fixed by assuming the same energy density for the distribution function and the field:
\begin{equation} \label{eq:energyParticlelization}
 \epsilon=T^{00}=\int \frac{\dd^3 \bvec{p}}{(2 \pi)^3} \ E \
    f_\sigma(\bvec x, \bvec p, t) \equiv E \left( \sigma(\bvec{x}),
   \dot{\sigma}(\bvec{x}) \right) 
\end{equation}
The chosen distribution function and particle density has to be
consistent with the decay process. Additionally, any momentum of the
field with $\bvec p \neq \bvec{0}$ implies a non-zero collective
velocity of the coarse-grained distribution function $f_\sigma$. This
effect is covered by a relativistic boost of the distribution function
with the mean velocity of the field via the four-velocity
$u^{\mu}= {p^{\mu}}/{E}$. The field energy and momentum, $E$ and
$\bvec{p}$, contained in the cell around $\bvec{x}$ are determined
according to (\ref{theory:fieldParticle:fieldEnergy}). In our
calculations we have used the boosted Boltzmann distribution for the
particles $q$, anti-particles $\bar q$ and the $\sigma$-particles,
\begin{equation}
  f_\sigma \sim \exp \left( - \frac{p_{\mu} \cdot u^{\mu}(\bvec x)}{T}\right) \ ,
\end{equation}
but the local energy relation (\ref{eq:energyParticlelization}) can
easily be extended to other distributions like the Bose-Einstein
distribution. After calculating a distribution function, $\sigma$
particles are sampled from $f_\sigma$ with Monte-Carlo methods. For
every sampled particle the decay probability is calculated.  In the
center of mass system of the particle, the decay probability is given by
\begin{equation}
  \Gamma_\sigma = \frac{g^2}{8 \pi m_\sigma} \sqrt{1 - \frac{4 m_q^2}{m_\sigma^2}}
\end{equation}
with a $\Gamma$ consistent with (\ref{eq:BreitWignerCrossSection}). If a
$\sigma$-particle decays, the energy and momentum of the particle is
calculated with the assumption of all particles being scalar
\begin{equation}
  \Delta E = E_\sigma, \quad \Delta \bvec{P} = \bvec{P}_\sigma.
\end{equation}
The resulting amount of energy and momentum is removed again from the
field around the interaction point with help of the four energy and
momentum difference equations (\ref{eq:theory:FieldManipulationEnergy})
and (\ref{eq:theory:FieldManipulationMomentum}) and the Gaussian
parameterization (\ref{eq:parameterization:gauss3D}). Again, this
parameterization leads to a small interaction volume from which the
energy is dissipated.

To come back to our notation of an interaction-probability distribution,
we use the above discussed concepts to formulate an interaction
probability density per numerical cell
\begin{equation}
\begin{split}
  P (\Delta E, &\Delta \bvec{P}, \Delta t ) =  \\
  \quad \quad & \sum^{N_{\textrm{cell}}}_{i,j} \delta \left( \Delta E -
    \sqrt{s}\right) \delta \left(\Delta \bvec P - \left (\bvec p_i +
      \bvec p_j \right) \right)
  \frac{\hat \sigma_{\bar q q \to \sigma} v_{\textrm{rel}}(s) \Delta t}{\Delta V N_{\textrm{test}}} \\
  & +\delta \left( \Delta E - E_\sigma \right) \delta \left( \Delta
    \bvec P - \bvec P_\sigma \right)
  \frac{\Gamma_\sigma(m_\sigma) \ n_\sigma(\phi(\bvec x), t) \Delta t}{\Delta V} \\
  & + \Pr_0 \delta \left( \Delta E \right) \delta \left( \Delta \bvec P
  \right)
\end{split}
\end{equation}
with the sum over all particles $N_{\textrm{cell}}$ in a cell.

\begin{figure}
\centering
\includegraphics[height=0.3\textheight, width=0.7\textwidth, keepaspectratio]{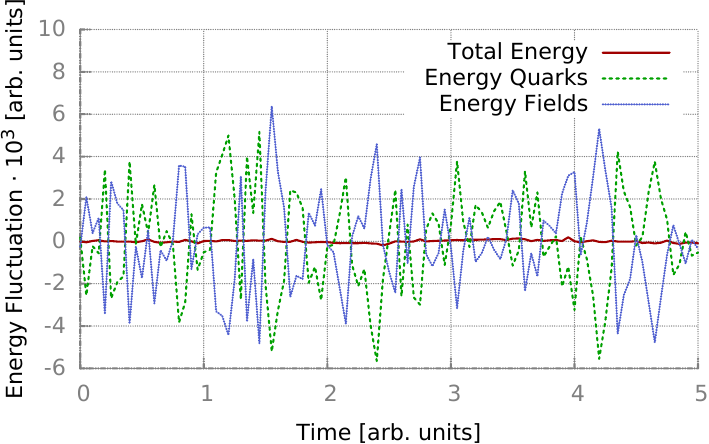}
\caption{(Color online) Energy fluctuations of the scalar field and the
  particles, $E(t) - \langle E \rangle$. Field and particles exchange
  energy by particle production and annihilation processes. While the
  total energy of the system is conserved and shows only numerical
  fluctuations, the energy of the components show thermal fluctuations,
  which are anticorrelated due to total-energy conservation.
  The relative fluctuations of the field's energy is $\sim 10^{-2}$, of the quarks $\sim 10^{-3}$. The total energy fluctuates on a scale of $|\Delta E|/\langle E \rangle \lesssim 5 \cdot 10^{-5}$.
  }
\label{fig:equilibriumBox:TotalEnergy}
\end{figure}

\begin{figure}
\centering
\includegraphics[width=0.6\textwidth,
keepaspectratio]{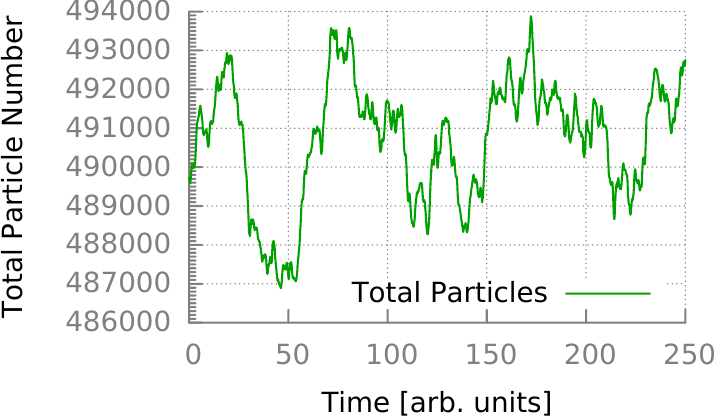}
\caption{(Color online) Total particle number in the thermal-box simulation. Particles can annihilate, their momentum and energy is transferred to the scalar
  field in form of scalar-field excitations. Because of the dynamic
  nature of this process, the total particle number fluctuates
  around the average thermal value. 
\label{fig:equilibriumBox:ParticleNumber}
}
\end{figure}

\begin{figure}
\centering
\includegraphics[width=0.6\textwidth, keepaspectratio]{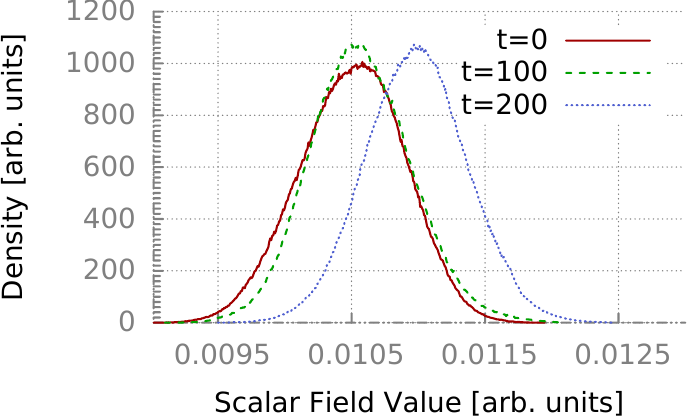}
\caption{(Color online) Distribution of the scalar field, showing the
  expected thermal Gauss-distributed fluctuations. The mean of the
  Gaussian can drift slowly with time due to fluctuations of the scalar
  mean field.  Particle annihilation increases the local fluctuations of
  the scalar field, and particle production damps them by dissipating
  energy from the interaction region.}
\label{fig:equilibriumBox:FieldDistribution}
\end{figure}

\begin{figure}
\centering
\includegraphics[width=0.3\textwidth,
keepaspectratio]{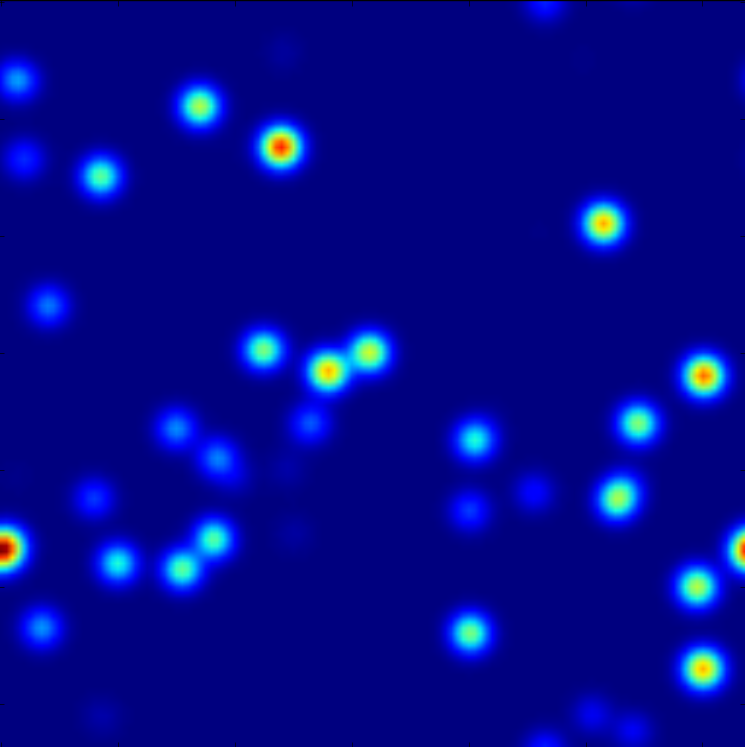}
\includegraphics[width=0.3\textwidth,
keepaspectratio]{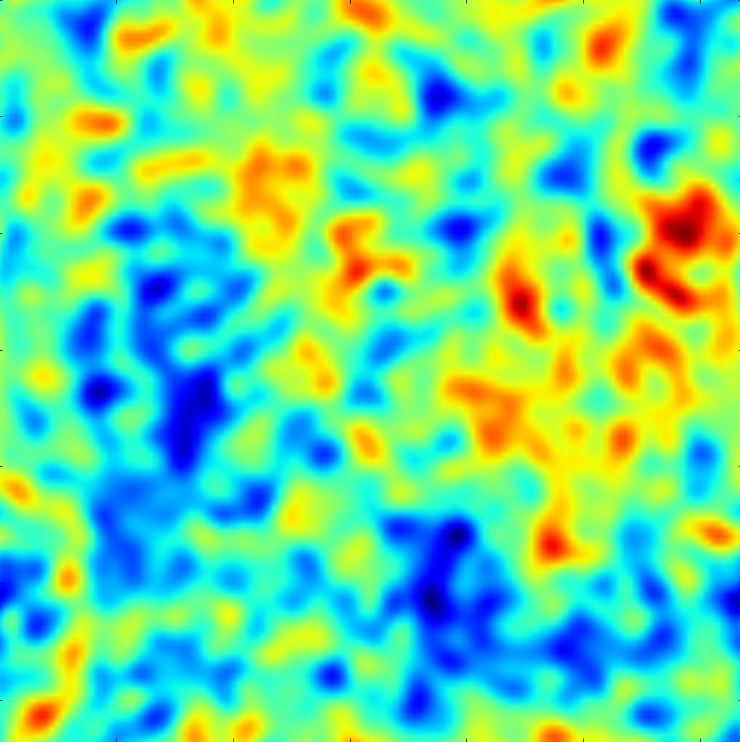}
\caption{(Color online) Plot of the 3D field with a cut in the $x-y$ plane at different
  times in the simulation.  The simulation starts with a uniform scalar
  field without any excited modes.  Due to particle creation and
  annihilation, field fluctuations are created dynamically within the
  simulation.  The color coding shows the value of the scalar field.
  Left: Some particles have annihilated and have created small,
  local excitations of the scalar field in form of moving, Gaussian
  shaped blobs. Right: The equilibrated field in the long-time limit: Due to the particle-field
  interactions, Gaussian fluctuations are dynamically created by microscopic interactions.
\label{fig:equilibriumBox:fluctuationsCreation}
}
\end{figure}

\begin{figure}
\centering
\includegraphics[width=0.6\textwidth,
keepaspectratio]{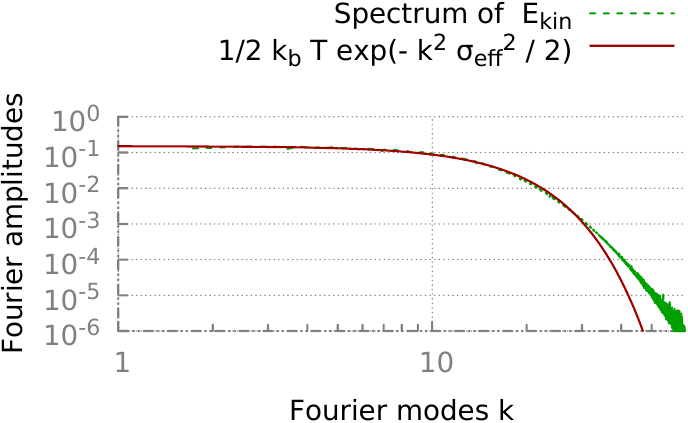}
\caption{(Color online) Spectrum of the field's kinetic energy
  $\dot \sigma^2/2$: While a thermal spectrum of a classical field has
  an average energy of $k T$ for every mode due to the equipartion
  theorem, with $T=0.15 \, \mathrm{arb. units}$, the simulated field has an
  evenly distributed energy per mode only for large wavelengths.  For
  small wavelengths, the spectrum is suppressed by a soft cut-off. This
  cut-off is due to the finte spatial extent of the interactions on the
  field. The deviation from the analytical result at higher modes can be
  explained by the non-linear potential in the field equations of
  motion. \label{fig:equilibriumBox:spectrumKineticEnergy} }
\end{figure}

\begin{figure}
\centering
\includegraphics[width=0.6\textwidth, keepaspectratio]{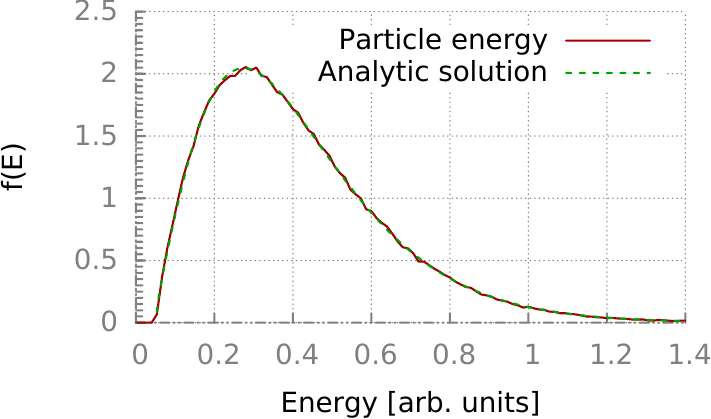}
\caption{(Color online) Distribution function of the particle energies
  showing the expected thermal Boltzmann distribution
  $f_q(E) \sim \exp \left(-E/T \right)$. By microscopic interactions
  particles and field can exchange energy and momentum and equilibrate
  to the same temperature $T=0.15 \textrm{arb. units}$ via annihilation and
  creation processes, demonstrating the proper implementation of the
  principle of detailed balance in our
  simulation. \label{fig:equilibriumBox:spectrumQuarks} }
\end{figure}

Figs.\ \ref{fig:equilibriumBox:TotalEnergy}-
\ref{fig:equilibriumBox:spectrumQuarks} show the results of the
described simulation. Over the whole simulation time, the total energy
and momentum stays constant, cf.\
Fig. \ref{fig:equilibriumBox:TotalEnergy}, while the particle number and
the scalar mean field show thermal fluctuations around a stable mean
value. Fig.\ \ref{fig:equilibriumBox:ParticleNumber} shows the total
particle number which fluctuates around a mean value. Field and
particles are coupled via chemical processes and exchange energy and
momentum, leading to fluctuations in both local energy densities and
their total energy, as shown in Fig.\
\ref{fig:equilibriumBox:TotalEnergy}. Any particle-annihilation process
increases the energy of the field, any decay of field modes leads to an
increase of the particle number. All fluctuations caused by these
interactions lead to a strong (negative) correlation between the
particle number and the mean field.  Overall, the total energy of the
system is conserved and shows only numerical fluctuations.

Fig.\ \ref{fig:equilibriumBox:FieldDistribution} represents the
scalar-field distribution, showing a stable Gaussian distribution.  For
a thermal field this is the expected behavior, similar to the Langevin
result (\ref{eq:dissipationFluctuationOscillator}). The width of the
Gaussian shows small thermal fluctuations over time, additionally does
the mean-value drift as a global fluctuation.

This is a remarkable result, because our scalar field shows the
distribution of a thermal field which usually is archived by coupling
the field to a heatbath via a Langevin-like process. A classical field,
coupled to Gaussian noise, would show an equipartition of energy of
$\frac{1}{2} k_B T$ for every degree of freedom.  However, integrating
over all modes leads to the famous classical ultraviolet catastrophe. In
our case, the field shows a slightly different distribution of energy
over the modes as shown in figure
\ref{fig:equilibriumBox:spectrumKineticEnergy}. For large wavelengths,
the modes are populated according to the equipartition theorem; for
small wavelengths the modes are strongly suppressed by the soft cutoff
given by (\ref{eq:cuttofSigma}) and (\ref{eq:cuttofSigma2}). The finite
extension of the interaction volume leads to a smeared distribution of
energy which is transferred to different modes. However, no modes with a
wavelength much smaller than the extension of the interaction volume can
be excited. The shape of the interaction parameterization
(\ref{eq:parameterization:gauss3D}) is directly reflected in the
spectrum of the kinetic energy, as can be seen by the analytic line in
the figure. The effective extension of the interaction volume
$\sigma_{\textrm{eff}}$ in Fig.\
\ref{fig:equilibriumBox:spectrumKineticEnergy} is given by
\begin{equation}
\sigma_{\textrm{eff}} = \sqrt{\sigma_x^2 + \sigma_y^2 + \sigma_z^2}.
\end{equation}
This soft cutoff is the first interesting feature of the proposed
method. The second aspect is that, while classical Langevin models
require an ad-hoc description of a stochastic force for thermalization,
the proposed method does not need a non-deterministic random source.
Instead, the random process is determined by a physically motivated
microscopic model for the interactions of the particles and fields,
having full control over the interaction rate as well as the energy and
momentum exchange. Fig.\ \ref{fig:equilibriumBox:fluctuationsCreation}
shows an initially vanishing scalar field with some field excitations
generated by particle interactions. These particle-annihilation
processes increase the field's energy, and large field modes are
created. After some time, these modes overlap and the field will start
to show a random, Gaussian distribution. Fig.\
\ref{fig:equilibriumBox:FieldDistribution} shows the field's
distribution after thermalization.  The distribution function of the
particles show a thermal Boltzmann distribution
$f_q(E) \sim \exp \left(-E/T \right)$ as shown in Fig.\
\ref{fig:equilibriumBox:spectrumQuarks}.  The temperature of the
particles is the same as the one of the thermal field fluctuations,
demonstrating the accurate implementation of the principle of detailed
balance in our numerical simulation.



\section{\label{sec:Applications}Possible Applications}

\begin{figure}
\centering
\includegraphics[width=0.7\textwidth,
keepaspectratio]{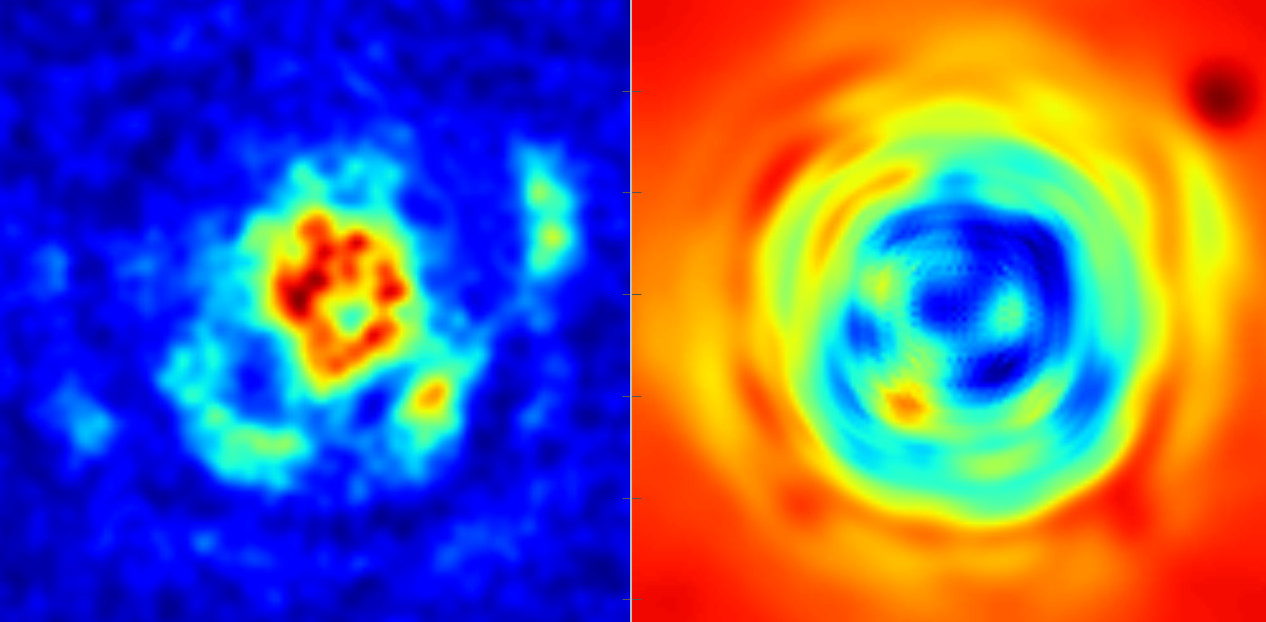}
\caption{(Color online) Rapid expansion of a hot quark-droplet as a
simply non-equilibrium setup for a heavy-ion collision. The interaction
between quarks and fields are implemented with the method purposed in
this paper. The chiral $\sigma$-field can remit latent energy, which is
freed by the phase transition, by creating particles. Results of this
study will be presented in an upcoming paper. Left: quark
density; Right: mean field value of the $\sigma$-field.
\label{fig:outlook:thermalBlobb}}
\end{figure}

We have applied the particle-field method to a transport simulation for
a model with scalar fields and particles. This method is now applied to
a linear $\sigma$-model \cite{WespPhDThesis} with the goal to
investigate the dynamics of the model near and at the phase transition,
where fluctuations are mainly driven by particle-field
interactions. Additionally the impact of observables of the order of the
phase transition is investigated, results of this study will be
presented in a forthcoming paper \cite{wespForthcoming}.  Figure
\ref{fig:outlook:thermalBlobb} shows the expansion of a hot
quark-droplet, which is a simple non-equilibrium setup for a heavy-ion
collision.  The Yukawa-coupling of the quarks to the chiral
$\sigma$-field determine their effective mass, while the expansion and
interactions drive fluctuations of the system. The coupling strength
determines the order of the phase transition, which has a strong impact
on the particle interactions, production and type of fluctuations.
Results of this study will be presented in a forthcoming paper
\cite{wespForthcoming}.

The method can easily be applied to other transport problems in
heavy-ion and nuclear physics, for example in heavy-quark simulations
with a hydrodynamical background \cite{PhysRevC.79.054907}. Besides the
already shown examples, the particle-field method can be employed
whenever noncontinuous interactions between particles and fields or
fields of different types have to be modeled.

In cosmology, particle creation within the inflationary phase of the big
bang are modeled by the dissipative part of an expanding scalar field
$\dot \phi$ \cite{Abbott198229, PhysRevD.51.4419}. Our method can be
used for numerical simulation of the initial fluctuations on large
scales. Another application in astro-physics could be to simulate pair
production in electric and magnetic fields near pulsars
\cite{PhysRevD.14.340} or to investigate the Sunyaev-Zel'dovich effect
induced by inverse Compton scattering \cite{0004-637X-523-1-78}. The
simulated Compton effect plays also an important role in plasma physics,
where damping of electromagnetic waves is affected by this process
induced by density fluctuations \cite{PhysRevA.46.1091}. In geophysics
earthquake crack- and wave-propagation are modeled with partial
differential equations \cite{1980565, Bao199885}. Our method could be
used to simulate local instabilities, which deploy energy within the
earthquake.

\section{Summary}

In this paper we have discussed various methods to simulate interactions
between particles and fields. Most of these methods are plagued by the
problem to allow only continuous changes of the fields, which are hard
to implement on the particle side or have some drawbacks like a
non-consistent physical interpretation or create an artificial
interaction-time scale. Additionally, methods like the Langevin equation
are effective theories which integrate out microscopic interactions, or
energy and momentum are not conserved.

In this paper we have presented a new method, which allows to describe
and simulate noncontinuous interactions with exact energy and momentum
conservation at all times.  The position, strength and time of the
interactions can be derived microscopically and simulated with
Monte-Carlo methods. To give examples for this method, we have applied
this method to a simple harmonic oscillator which dissipates energy by
losing discrete energy quanta.  In a second example the harmonic
oscillator is additionally coupled to a stochastic force, the overall
system behaves like an oscillator coupled to white-noise fluctuating
forces as in a classical Langevin-equation.  This example is generalized
to a $(1+1)$D scalar field. All three systems behave in very good
agreement with the analytic expectations. In the last and most complex
example a $(3+1)$D scalar field is coupled to an ensemble of particles
and anti-particles.  Excitations of the scalar field are treated as
particles which can decay via microscopic interactions and create new
particle-anti-particle pairs.  Additionally particles can annihilate and
create field quanta. By obeying the principle of detailed balance both
particles and fields reach thermal equilibrium with the same
temperature, fluctuations on the field are generated dynamically by
field-particle interactions and have not to be implemented via ad-hoc
assumptions like Gaussian white noise.

In an outlook several potential interesting applications of various
kinds of simulations have been suggested.

\section*{Acknowledgments}

This work was partially supported by the Bundesministerium f{\"u}r
Bildung und Forschung (BMBF F{\"o}rderkennzeichen 05P12RFFTS) and by the
Helmholtz International Center for FAIR (HIC for FAIR) within the
framework of the LOEWE program (Landesoffensive zur Entwicklung
Wissenschaftlich-{\"O}konomischer Exzellenz) launched by the State of
Hesse.  C.\ W.\ and A.\ M.\ acknowledge support by the Helmholtz
Graduate School for Hadron and Ion Research (HGS-HIRe), and the
Helmholtz Research School for Quark Matter Studies in Heavy Ion
Collisions (HQM). Numerical computations have been performed at the
Center for Scientific Computing (CSC).  H.\ v.\ H.\ has been supported
by the Deutsche Forschungsgemeinschaft (DFG) under grant number GR
1536/8-1.  C.W. has been supported by BMBF under grant number 0512RFFTS.

\input{appendix.tex}

\bibliography{Bibliography}

\end{document}

%% file: appendix.tex
\appendix

\section{The principle of detailed balance for particle decay and
  recombination}

In this appendix we briefly summarize our treatment of the particle-annihilation process and the corresponding inverse decay process with
regard to the principle of detailed balance. To keep the discussion at
the most simple level possible, we consider a purely bosonic model. The
relevant interaction part of the Lagrangian reads
\begin{equation}
\label{app.1}
\mathcal{L}_{\text{I}} = \lambda \sigma q^* q,
\end{equation}
where $\sigma$, describing $\sigma$ particles of mass $m_{\sigma}$, is a
real and $q$ a complex scalar field, describing particles with
mass $2 m<m_{\sigma}$. At tree level the matrix elements for decay and
recombination are the same and simply given by
\begin{alignat}{2}
\label{app.2.1}
\mathcal{M}_{\sigma \rightarrow q \bar{q}}
&= \parbox{2cm}{\includegraphics[width=\linewidth]{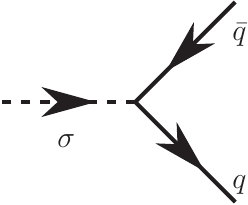}}=-\ii
  \lambda =\mathcal{M}, \\
\label{app.2.2}
\mathcal{M}_{q \bar{q} \rightarrow \sigma}
&= \parbox{2cm}{\includegraphics[width=\linewidth]{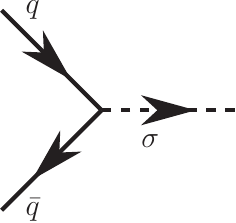}}=-\ii
  \lambda=\mathcal{M}.
\end{alignat}
Assuming net-$q$-particle neutrality ($f_{q}=f_{\bar{q}}$), the
corresponding Boltzmann-Vlasov equations for $q$ and $\sigma$ particles
thus read
\begin{equation}
\begin{split}
\label{app.3}
\partial_t f_q + \frac{\bvec{p}}{E_q} \cdot \frac{\partial f_q}{\partial
  \bvec{x}} + \bvec{F}(\bvec{x},\bvec{p},t) \frac{\partial f_q}{\partial
  \bvec{p}} &= \mathcal{C}_q[f_q,f_{\sigma}],\\
\partial_t f_{\sigma} + \frac{\bvec{q}}{E_{\sigma}} \cdot \frac{\partial f_{\sigma}}{\partial
  \bvec{x}} &= \mathcal{C}_{\sigma}[f_q,f_{\sigma}].
\end{split}
\end{equation}
The collision terms are given by the transition rates for creation and
annihilation of $q$ and $\sigma$ particles,
\begin{alignat}{2}
\begin{split}
\label{app.4.1}
\mathcal{C}_q = \frac{\lambda^2}{2 E_q} & \int_{\R^3}  \frac{\dd^3
  \bvec{p}_2}{(2
  \pi)^3 2 E_{q2}}  \int_{\R^3} \frac{\dd^3 \bvec{q}}{(2 \pi)^3 2 E_{\sigma}} \\
& (2 \pi)^4 \delta^{(4)}(p_1+p_2-q) \\
& \left [f_{\sigma}(\bvec{x},\bvec{q},t) - f_{q}(\bvec{x},\bvec{p},t)
  f_{\bar{q}}(\bvec{x},\bvec{p}_2,t) \right],
\end{split}
\\
\begin{split}
\label{app.4.2}
\mathcal{C}_{\sigma} = \frac{\lambda^2}{2 E_{\sigma}} &\int_{\R^3} 
\frac{\dd^3 \bvec{p}}{(2 \pi)^3 2 E_{q}}  \int_{\R^3} \frac{\dd^3 \bvec{p}_2}{(2
  \pi)^3 2 E_{q2}} \\
& (2 \pi)^4 \delta^{(4)}(p_1+p_2-q)\\
& \left [ f_{q}(\bvec{x},\bvec{p},t) f_{\bar{q}}(\bvec{x},\bvec{p}_2,t)
  - f_{\sigma}(\bvec{x},\bvec{q},t) \right].
\end{split}
\end{alignat}
These collision terms obviously fulfill the principle of detailed
balance. Due to Boltzmann's H theorem, (local) equilibrium is reached
when the square brackets in the above integrals vanish, and this is
indeed the case for the Boltzmann distributions,
\begin{equation}
\label{app.5.1}
f_q(p)=\exp(-p \cdot u/T), \quad f_{\sigma}(q) = \exp(-q \cdot u/T),
\end{equation}
where $u=u(t,\bvec{x})$ is the four-velocity flow field, and
$T=T(t,\bvec{x})$ the temperature field \cite{Cercignani:2002}. Further,
the four-momenta of all particles in (\ref{app.4.1}) and (\ref{app.4.2})
are assumed to fulfill the corresponding on-shell conditions, i.e.,
$E_q=\sqrt{\bvec{p}^2+m_q^2}$ and
$E_{\sigma}=\sqrt{\bvec{q}^2+m_{\sigma}^2}$. This causes the difficulty
that the recombination process (\ref{app.2.2}) is ineffective due to the
on-shell constraint $(p+p_2)^2=m_{\sigma}^2$, which is artificial since
an unstable particle has a finite-width spectral function. Thus, using
the width according to the tree-level matrix element (\ref{app.2.2}),
\begin{equation}
\label{app.5.2}
\Gamma_{\sigma}=\frac{\lambda^2}{8 \pi m_{\sigma}} \sqrt{1-\frac{4 m_q^2}{m_{\sigma}^2}},
\end{equation}
we substitute in (\ref{app.4.1}) the integration operator
\begin{equation}
\label{app.6}
\int_{\R^3} \frac{\dd^3 \bvec{q}}{(2 \pi)^3 2 E_{\sigma}} = \int_{\R^4}
\frac{\dd^4 q}{(2 \pi)^4} \Theta(q^0) \delta[(q^0)^2-E_{\sigma}^2]
\end{equation}
by
\begin{equation}
\label{app.7}
\int_{\R^4} \frac{\dd^4 q}{(2 \pi)^4} \frac{\Gamma}{(q_0-E_{\sigma})^2+\Gamma_{\sigma}^2/4}.
\end{equation}
For consistency and to preserve detailed balance, also (\ref{app.4.2}) has to be averaged by the
Breit-Wigner distribution in the same way, i.e., we have to substitute
\begin{equation}
\label{app.8}
\delta(q^0-p_1^0-p_2^0) \rightarrow \frac{1}{2 \pi}
\frac{\Gamma}{(p_1^0+p_2^0-E_{\sigma})^2+\Gamma_{\sigma}^2/4}
\end{equation}
for the energy-conserving $\delta$ distribution in the collision term. A
fully self-consistent treatment of particles of finite width within a
kinetic approach \cite{kv97,Ivanov:1999tj} is beyond the scope of this
paper.
